\documentclass[prd,letterpaper,twocolumn,preprintnumbers,nofootinbib]{revtex4}
\usepackage[letterpaper, hdivide={1.91cm,,1.165cm}, vdivide={1.83cm,,3.6cm}]{geometry}

\usepackage{slashed}
\usepackage{amsmath,amssymb}
\usepackage{graphicx}
\usepackage{units}
\usepackage{bbold}
\usepackage{xcolor}
\usepackage{dsfont}
\usepackage[hyperfootnotes=false,colorlinks,citecolor=blue]{hyperref}

\usepackage{comment}
\usepackage[normalem]{ulem}

\newcommand{\hc}{\ensuremath{\text{h.c.}}}
\newcommand{\BR}{\ensuremath{\text{BR}}}

\begin{document}

\title{Lepton flavor violation by two units}

\author{Julian Heeck}
\email[Email: ]{heeck@virginia.edu}
\thanks{ORCID: \href{https://orcid.org/0000-0003-2653-5962}{0000-0003-2653-5962}.}
\affiliation{Department of Physics, University of Virginia,
Charlottesville, Virginia 22904-4714, USA}

\author{Mikheil Sokhashvili}
\email[Email: ]{ms2guc@virginia.edu}
\thanks{ORCID: \href{https://orcid.org/0000-0003-0844-7563}{0000-0003-0844-7563}.}
\affiliation{Department of Physics, University of Virginia,
Charlottesville, Virginia 22904-4714, USA}

\hypersetup{
pdftitle={Lepton flavor violation by two units},   
pdfauthor={Julian Heeck, Mikheil Sokhashvili}
}

\begin{abstract}

Charged lepton flavor violation arises in the Standard Model Effective Field Theory at mass dimension six. The operators that induce neutrinoless muon and tauon decays are among the best constrained and are sensitive to new-physics scales up to $10^7\,$GeV. An entirely different class of lepton-flavor-violating operators violates lepton flavors by two units rather than one and does not lead to such clean signatures. Even the well-known case of muonium--anti-muonium conversion that falls into this category is only sensitive to two out of the three $\Delta L_\mu = - \Delta L_e = 2$ dimension-six operators. We derive constraints on many of these operators from lepton flavor universality and show how to make further progress with future searches at Belle II and future experiments such as $Z$ factories or muon colliders.

\end{abstract}

\maketitle

\section{Introduction}
\label{sec:intro}

The Standard Model (SM) has accidental symmetries that lead to the conservation of electron, muon, and tauon number~\cite{Calibbi:2017uvl,Davidson:2022jai}. Neutrino oscillations are proof that these symmetries are ultimately violated in nature, but might be unobservably suppressed by the tiny $m_\nu^2$ in the charged-lepton sector in the worst-case scenario~\cite{Heeck:2016xwg}. Luckily, many SM extensions violate these flavor symmetries and can lead to testable flavor-violating processes, unsuppressed by the neutrino mass~\cite{Calibbi:2017uvl,Davidson:2022jai}.

To stay model agnostic, we can resort to the Standard Model Effective Field Theory (SMEFT, see Ref.~\cite{Isidori:2023pyp} for a recent review), which extends the SM by higher-dimensional effective operators suppressed by powers of some high scale $\Lambda$~\cite{Weinberg:1979sa,Weinberg:1980bf}. Integrating out heavy new particles produces exactly these kind of operators.
Ordering the higher-dimensional operators by their mass dimension (i.e.~powers of $1/\Lambda$), the leading one is Weinberg's $LLHH/\Lambda$ that generates Majorana neutrino masses~\cite{Weinberg:1979sa}, solving one of the SM's biggest problems.
At $1/\Lambda^2$, or mass dimension $d=6$, there are thousands of operators~\cite{Buchmuller:1985jz,Grzadkowski:2010es}, most of which just lead to small corrections to processes that are already allowed in the SM. But some of them violate the SM's accidental symmetries and thus lead to completely different processes that in principle have zero background.

Dimension-six operators with $\Delta L_\alpha = -\Delta L_\beta = 1-\Delta L_\gamma = 1$, where $\alpha$, $\beta$, and $\gamma$ are distinct lepton flavors, can be probed in decays of an $\alpha$ or $\beta$ lepton, leading to fully visible neutrinoless two-body signatures such as~$\mu\to e\gamma$ or $\tau\to \mu \pi^0$. These are the most studied lepton-flavor-violating operators/signatures, both theoretically and experimentally, probing $\Lambda$ scales up to $\unit[10^7]{GeV}$ in the muon sector and $\unit[10^4]{GeV}$ in the tauon sector~\cite{Calibbi:2017uvl,Davidson:2022jai}.

Alas, there are 21 $d=6$ operators that violate lepton flavor by \emph{two} units rather than one, and thus might not give rise to neutrinoless decays. For example, operators of the form $\mu\mu \bar{e}\bar{e}$ violate $\Delta L_\mu = -\Delta L_e = 2$ and do not lead to on-shell muon decays. In this particular example, muonium--anti-muonium conversion provides a good alternative signature for two out of three $\mu\mu \bar{e}\bar{e}$ operators. However, once we have a look at the tauon sector, e.g.~$\tau\tau \bar{e}\bar{e}$, even the leptonium option is removed and the operators are seemingly unconstrained despite violating lepton flavor and being of low mass dimension.

Here, we will investigate such $\Delta L_\alpha = 2$ operators and identify possible ways to constrain them. 
Weinberg's $d=5$ operator already contains $\Delta L_\alpha = 2$ pieces, but since they are suppressed by neutrino masses these operators are rendered unobservable, with the possible exception of neutrinoless double beta decay~\cite{Rodejohann:2011mu}. 
At $d=6$, all $\Delta L_\alpha = 2$ operators are part of the four-lepton operators~\cite{Grzadkowski:2010es}
\begin{align}
\mathcal{L} & \supset \sum_{a,b,c,d = e,\mu,\tau} \left[ y^{LL}_{abcd}\bar{L}_a \gamma^\alpha L_b\, \bar{L}_c \gamma_\alpha L_d\right.  \\
&\left. \quad + y^{LR}_{abcd}\bar{L}_a \gamma^\alpha L_b\, \bar{\ell}_c\gamma_\alpha \ell_d  + y^{RR}_{abcd}\bar{\ell}_a \gamma^\alpha \ell_b\, \bar{\ell}_c \gamma_\alpha \ell_d\right]+ \hc\,,\nonumber
\end{align}
where $L_a$ is the left-handed lepton doublet of flavor $a$,  $\ell_a$ the right-handed charged lepton of flavor $a$, and the $y = \mathcal{O}(\Lambda^{-2})$ are the Wilson coefficients of mass dimension $-2$.
UV-complete realizations of these operators involve neutral or doubly-charged bosons and can for example be found in Refs.~\cite{Cuypers:1996ia,Altmannshofer:2016brv,Fukuyama:2021iyw,Goudelis:2023yni}.

We stress that the $\Delta L_\alpha=2$ operators under investigation here are fundamentally distinct from the more familiar $\Delta L_\alpha=1$ operators; since they carry different lepton numbers it is easily possible to impose a symmetry in the charged-lepton sector that would forbid $\Delta L_\alpha=1$ but allow for $\Delta L_\alpha=2$. Examples are $U(1)$ flavor symmetries or their $\mathbb{Z}_N$ subgroups~\cite{Heeck:2016xwg},  and lepton flavor triality $\mathbb{Z}_3$~\cite{Ma:2010gs,Cao:2011df,Bigaran:2022giz}, which arises in many neutrino-mass models based on discrete symmetries such as $A_4$~\cite{Altarelli:2005yx,Ishimori:2010au,Holthausen:2012wz} and only allows for operators of the form $\bar{\tau}\bar{\mu} ee$, $\bar{\tau}\bar{e} \mu\mu$, and $\bar{\tau}\bar{\tau} e\mu$.
 Renormalization-group running can turn $\Delta L_\alpha=1$ operators into $\Delta L_\alpha=2$, but not vice versa, rendering dedicated searches for $\Delta L_\alpha=2$ absolutely necessary to cover these blind spots.

\section{\texorpdfstring{$\bar{\mu}\bar{\mu} e e$}{mu-mu-e-e}}

\begin{table*}[th]
    \centering
\begin{tabular}{ l|l|l|l }
 \hline
 Wilson coefficient& Upper limit & Process & Violated quantum numbers \\
 \hline \hline
     $|y_{\mu e \mu e}^{LL} + y_{\mu e \mu e}^{RR}|$ & $( \unit[3.2]{TeV})^{-2}$ ~[$90\%\,$C.L.] & Mu-to-$\overline{\text{Mu}}$ ~\cite{Willmann:1998gd} & $\Delta L_\mu =- \Delta L_e = 2$\\
 \hline
     $|y_{\mu e \mu e}^{LR}|$ & $( \unit[3.8]{TeV})^{-2}$ ~[$90\%\,$C.L.] & Mu-to-$\overline{\text{Mu}}$ ~\cite{Willmann:1998gd} & $\Delta L_\mu =- \Delta L_e = 2$\\
 \hline
     $|y_{\mu e \mu e}^{LL} - y_{\mu e \mu e}^{RR}|$ & $( \unit[0.74]{TeV})^{-2}$ ~[$95\%\,$C.L.] & $\Gamma(\mu \to e \nu \bar\nu)/\Gamma(\tau \to \mu \nu \bar\nu)$ ~\cite{HeavyFlavorAveragingGroup:2022wzx} & $\Delta L_\mu =- \Delta L_e = 2$\\
 \hline \hline
     $|2y_{\tau e \tau e}^{LL}|, |y_{\tau e \tau e}^{LR}|$ & $( \unit[0.67]{TeV})^{-2}$ ~[$95\%\,$C.L.] & $\Gamma( \tau \to e \nu \bar\nu) /\Gamma(\tau \to \mu \nu \bar\nu)$ ~\cite{HeavyFlavorAveragingGroup:2022wzx}~\cite{Belle_TAU2023} & $\Delta L_\tau =- \Delta L_e = 2$\\
 \hline 
     $|y_{\tau e \tau e}^{RR}|$ & $( \unit[1.2]{GeV})^{-2}$ ~[$95\%\,$C.L.] & $Z \to \tau^\pm\tau^\pm e^\mp e^\mp$ & $\Delta L_\tau =- \Delta L_e = 2$\\
 \hline \hline
     $|2y_{\tau \mu \tau \mu}^{LL}|, |y_{\tau \mu \tau \mu}^{LR}|$ & $( \unit[0.63]{TeV})^{-2}$ ~[$95\%\,$C.L.] & $\Gamma(\tau \to \mu \nu \bar\nu)/\Gamma(\tau \to e \nu \bar\nu)$ ~\cite{HeavyFlavorAveragingGroup:2022wzx}~\cite{Belle_TAU2023} & $\Delta L_\tau =- \Delta L_\mu = 2$\\
 \hline
     $|y_{\tau \mu \tau \mu}^{RR}|$ & $( \unit[1.2]{GeV})^{-2}$ ~[$95\%\,$C.L.] & $Z \to \tau^\pm\tau^\pm \mu^\mp \mu^\mp$ & $\Delta L_\tau =- \Delta L_\mu = 2$\\
 \hline \hline
     $|y_{e \tau \mu \tau}^{LL}|, |y_{\mu \tau e \tau}^{LR}|$ & $( \unit[0.60]{TeV})^{-2}$ ~[$95\%\,$C.L.] & $\Gamma(\tau \to e \nu \bar\nu)/\Gamma(\mu \to e \nu \bar\nu)$ ~\cite{HeavyFlavorAveragingGroup:2022wzx} & $\Delta L_\tau= -2 \Delta L_\mu =-2 \Delta L_e = 2$\\
 \hline
      $|y_{e \tau \mu \tau}^{LR}|$ & $( \unit[0.55]{TeV})^{-2}$ ~[$95\%\,$C.L.] & $\Gamma(\tau \to e \nu \bar\nu)/\Gamma(\tau \to \mu \nu \bar\nu)$ ~\cite{HeavyFlavorAveragingGroup:2022wzx}~\cite{Belle_TAU2023} & $\Delta L_\tau= -2 \Delta L_\mu =-2 \Delta L_e = 2$\\
 \hline
     $|y_{e \tau \mu \tau}^{RR}|$ & $( \unit[1]{GeV})^{-2}$ ~[$95\%\,$C.L.] & $Z \to \tau^\pm\tau^\pm e^\mp \mu^\mp$ & $\Delta L_\tau= -2 \Delta L_\mu =-2 \Delta L_e = 2$\\
 \hline \hline
     $|y_{\mu e \tau e}^{LL}|, |y_{\mu e \tau e}^{LR}|, |y_{\tau e \mu e}^{LR}|, |y_{\mu e \tau e}^{RR}|$ & $(\unit[10]{TeV})^{-2}$ ~[$90\%\,$C.L.] & $\tau \to \bar{\mu} ee$ ~\cite{Hayasaka:2010np} & $\Delta L_e= -2 \Delta L_\tau =-2 \Delta L_\mu = 2$\\
 \hline \hline
     $|y_{e \mu \tau \mu}^{LL}|, |y_{e \mu \tau \mu}^{LR}|, |y_{\tau \mu e  \mu}^{LR}|, |y_{e \mu \tau \mu}^{RR}|$ & $(\unit[8.8]{TeV})^{-2}$ ~[$90\%\,$C.L.] & $\tau \to \bar{e} \mu \mu$ ~\cite{Hayasaka:2010np} & $\Delta L_\mu=-2 \Delta L_\tau= -2 \Delta L_e  = 2$\\
 \hline
\end{tabular}
    \caption{Current limits on the magnitudes of the 21 $\Delta L_\alpha = 2$ dimension-six Wilson coefficients as well as the corresponding processes and the violated quantum numbers. Details are given in the text.
    }
    \label{tab:limits}
\end{table*}

Let us first focus on the well-known $d=6$ operators with $\Delta L_\mu =- \Delta L_e = 2$:
\begin{align}
\begin{split}
\mathcal{L}& \supset y^{LL}_{\mu e\mu e}\bar{L}_\mu \gamma^\alpha L_e\, \bar{L}_\mu \gamma_\alpha L_e+ y^{LR}_{\mu e\mu e}\bar{L}_\mu \gamma^\alpha L_e\, \bar{\ell}_\mu \gamma_\alpha \ell_e \\
&\quad + y^{RR}_{\mu e\mu e}\bar{\ell}_\mu \gamma^\alpha \ell_e\, \bar{\ell}_\mu \gamma_\alpha \ell_e+ \hc
\end{split}
\label{eq:mumuee}
\end{align}
All three operators will contribute to muonium--antimuonium conversion~\cite{Conlin:2020veq,Fukuyama:2021iyw,Petrov:2022wau,Fukuyama:2022fwi,Fukuyama:2023drl}. Using the experimental setup of the PSI experiment that provides the strongest limit to date~\cite{Willmann:1998gd}, the conversion probability $P$ takes the approximate form~\cite{Fukuyama:2021iyw}
\begin{align}
\begin{split}
P &\simeq \frac{7.58\times 10^{-7}}{G_F^2} |y^{LL}_{\mu e \mu e} + y^{RR}_{\mu e \mu e} - 1.68 y^{LR}_{\mu e \mu e}|^2  \\
 &\quad + \frac{4.27\times 10^{-7}}{G_F^2} |y^{LL}_{\mu e \mu e} + y^{RR}_{\mu e \mu e} + 0.68 y^{LR}_{\mu e \mu e}|^2 \,.
\end{split}
\end{align}
The PSI limit $P< 8.3\times 10^{-11}$~\cite{Willmann:1998gd}  at $90\%\,$C.L. 
then puts strong upper limits of order $(\unit[3]{TeV})^{-2}$ on the Wilson coefficients $|y^{LL}_{\mu e \mu e} + y^{RR}_{\mu e \mu e}|$ and $|y^{LR}_{\mu e \mu e}|$, but is insensitive to the linear combination $y^{LL}_{\mu e \mu e} - y^{RR}_{\mu e \mu e}$, which corresponds to the vector--axial-vector operator $\bar{\mu}\gamma_\alpha e\, \bar{\mu}\gamma^\alpha \gamma_5 e$. 
Keeping the other Wilson coefficients free, we obtain the limits $|y^{LL}_{\mu e \mu e} + y^{RR}_{\mu e \mu e}| < (\unit[2.9]{TeV})^{-2}$ and $|y^{LR}_{\mu e \mu e}| < (\unit[3.4]{TeV})^{-2}$. 
We can put stronger limits, shown in Tab.~\ref{tab:limits}, by setting one of the linear combinations to zero, i.e.~forbidding interference between the contributions.
Ref.~\cite{Conlin:2020veq} has shown that muonium conversion is actually also affected by $y^{LL}_{\mu e \mu e} - y^{RR}_{\mu e \mu e}$  through the muonium widths difference, but since these effects are further suppressed by the Fermi constant $G_F$, the resulting limits are only of order $(\unit[1.2]{GeV})^{-2}$, probing scales far \emph{below} the electroweak scale and thus not particularly relevant for the SMEFT. 
Future experiments such as the Muonium-to-Antimuonium Conversion Experiment (MACE) at the China Spallation Neutron Source (CSNS)~\cite{Bai:2022sxq} and  a new setup at the Japan Proton Accelerator Research Complex (J-PARC)~\cite{Kawamura:2021lqk} are expected to improve the old PSI bounds by orders of magnitude.

\begin{figure}[tb]
    \centering
    \includegraphics[width=0.35\textwidth]{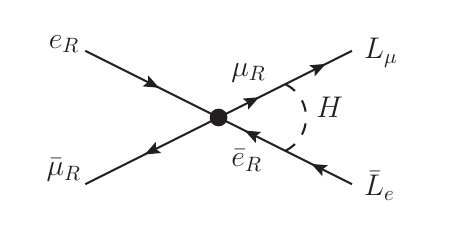}
    \caption{Example Feynman diagram that shows the conversion of an $ \bar{\ell}_\mu \gamma^\alpha \ell_e\, \bar{\ell}_\mu \gamma_\alpha \ell_e$ operator (filled circle) to a $ \bar{L}_\mu \gamma^\alpha L_e\, \bar{\ell}_\mu \gamma_\alpha \ell_e $ operator at loop level using the SM Higgs interactions.
    }
    \label{fig:muemue_mixing}
\end{figure}

Note that the three operators in Eq.~\eqref{eq:mumuee} carry the same quantum numbers and thus mix via loops or renormalization group equations~\cite{Jenkins:2013wua}. In the SMEFT energy region above the electroweak scale, this requires insertions of the lepton Yukawa couplings $y_\ell \equiv m_\ell/\unit[174]{GeV}$, see Fig.~\ref{fig:muemue_mixing}. For example, the operator with coefficient $y^{LL}_{\mu e \mu e} - y^{RR}_{\mu e \mu e}$ then generates a $y^{LR}_{\mu e \mu e}$ operator of magnitude
\begin{align}
y^{LR}_{\mu e \mu e} \simeq \frac{y_e y_\mu }{16\pi^2}\left(y^{LL}_{\mu e \mu e} - y^{RR}_{\mu e \mu e}\right) .
\label{eq:loop}
\end{align}
The tiny prefactor $y_e y_\mu /16\pi^2\simeq 10^{-11}$ renders these effects small and gives irrelevant limits on $y^{LL}_{\mu e \mu e} - y^{RR}_{\mu e \mu e}$.

Better limits on the linear combination $y^{LL}_{\mu e \mu e}-y^{RR}_{\mu e \mu e}$ can be obtained by noticing that the underlying operator contains both neutrinos and charged leptons and thus leads to the muon decay $\mu^-\to e^- \bar\nu_\mu\nu_e$. Since we are looking at $\Delta L_\mu = -\Delta L_e = 2$ processes, this decay does not interfere with the $\Delta L_\mu = \Delta L_e = 0$ SM decay $\mu^-\to e^- \nu_\mu\bar\nu_e$. We find that our operators from Eq.~\eqref{eq:mumuee} generate exactly the same electron energy spectrum as the SM decay, so the Michel spectrum remains unperturbed; only the overall muon lifetime or decay rate is affected:
\begin{align}
    \Gamma_\mu = \Gamma_\mu^\text{SM} \left(1 + \frac{|2y_{\mu e \mu e}^{LL}|^2 + |y_{\mu e \mu e}^{LR}|^2}{(2\sqrt{2} G_F)^2} \right) ,
\end{align}
even including radiative QED corrections.
Since $|y_{\mu e \mu e}^{LR}|$ is already constrained to be tiny from the muonium limits, we can safely neglect it here.
To obtain limits on the Wilson coefficients, we employ lepton-flavor-universality tests~\cite{Fujii:1993su},\footnote{Alternatively, one could track the impact of the so-modified Fermi constant on electroweak precision data~\cite{Bigaran:2022giz}, which is however currently difficult given the recent anomalous $W$-mass measurement by CDF-II~\cite{CDF:2022hxs}.} i.e.~we calculate  $\Gamma_\mu/\Gamma_{\tau \to e \nu \nu} $,
\begin{align}
\begin{split}
    &|2y_{\mu e \mu e}^{LL}|^2  =8 G_F^2 \left( \frac{\Gamma_{\mu \to e \nu \nu}^\text{exp}  m_\tau^5 f \left[ \frac{m_e^2}{m_\tau^2} \right] R_W^{e\tau} R^\tau_\gamma } {\Gamma_{\tau \to e \nu \nu}^\text{exp} m_\mu^5 f \left[ \frac{m_e^2}{m_\mu^2} \right] R_W^{e\mu} R^\mu_\gamma } - 1 \right),
\end{split}
\end{align}
using the definitions and experimental values from Ref.~\cite{HeavyFlavorAveragingGroup:2022wzx},
which gives
\begin{align}
    |2y_{\mu e \mu e}^{LL}|^2 = \left(-2.42 \pm 3.10\right) \unit{TeV^{-4}}.
\end{align}
Using  instead the ratio $\Gamma_\mu/\Gamma^\text{SM} (\tau \to \mu\nu\nu)$ puts a better limit,
\begin{align}
    |2y_{\mu e \mu e}^{LL}|^2 = \left(-6.10 \pm 3.13\right) \unit{TeV^{-4}} \,,
    \label{eq:LFUVlimitmuemue}
\end{align}
because the two-decades old measured $\tau \to \mu\nu\nu$ rate shows a $2\sigma$ increase with respect to the SM. This is likely a statistical fluctuation or systematic effect since a recent preliminary measurement of $\Gamma(\tau \to \mu\nu\nu)/\Gamma(\tau \to e\nu\nu)$ at Belle II (combined with older measurements) is perfectly compatible with the SM~\cite{Belle_TAU2023}:
\begin{align}
    \frac{\Gamma(\tau \to \mu\nu\nu)}{\Gamma(\tau \to e\nu\nu)} = \left(1.0009\pm 0.0027\right)     \frac{\Gamma(\tau \to \mu\nu\nu)_\text{SM}}{\Gamma(\tau \to e\nu\nu)_\text{SM}}\,.
    \label{eq:belleII}
\end{align}
The difference in the limits from $\tau\to e\nu\nu$ and  $\tau\to \mu\nu\nu$ will therefore probably decrease with dedicated measurements of the tauon branching ratios at Belle II. In the meantime, we construct one-sided $95\%$~C.L.~confidence intervals from the above  to find a 
limit on $|2y_{\mu e \mu e}^{LL}|$ of $1.8/\unit{TeV^2}$, which can also be written as 
\begin{align}
    |(y_{\mu e \mu e}^{LL} - y_{\mu e \mu e}^{RR}) + (y_{\mu e \mu e}^{LL} + y_{\mu e \mu e}^{RR})| < 1.8/\unit{TeV^2}. 
\end{align}
As $y_{\mu e \mu e}^{LL} + y_{\mu e \mu e}^{LL}$ is already constrained to be $\sim 20$ times smaller than the first term we obtain an upper bound on the Wilson-coefficient linear combination that is unconstrained by the muonium of $|y_{\mu e \mu e}^{LL} -y_{\mu e \mu e}^{RR}| < 1.8/\unit{TeV^2}$ (Tab.~\ref{tab:limits}).
This is the strongest limit on the remaining Wilson coefficients, corresponding to new-physics scales of $\unit[0.74]{TeV}$, above the electroweak scale and thus perfectly applicable to our SMEFT ansatz.
Since the uncertainties on the flavor-universality ratios are dominated by the tau lifetime and branching ratios, these are the quantities that need to be measured more precisely in order to improve the bound on $|y_{\mu e \mu e}^{LL} -y_{\mu e \mu e}^{RR}|$; Belle II will likely achieve this in the near future~\cite{Belle_TAU2023}. 

Although not currently relevant, let us mention some other experiments and signatures that could play a future role in constraining the $y_{\mu e\mu e}$ Wilson coefficients.  
$\bar{\mu}\bar{\mu}ee$ operators involving neutrinos can induce mixed-flavor neutrino-trident effects, e.g.~$\nu_\mu X\to \nu_e \mu^+ e^- X$ or $\nu_e X\to \nu_\mu e^+ \mu^- X$, which could be probed in future neutrino detectors such as DUNE~\cite{Magill:2016hgc,Magill:2017mps,Ballett:2018uuc,Altmannshofer:2019zhy}. Due to the non-interference with SM amplitudes the effects are expected to be small, roughly
\begin{align}
    \frac{\sigma (\nu_\mu X\to \nu_e \mu^+ e^- X) }{\sigma (\nu_\mu X\to \nu_e \mu^- e^+ X)_\text{SM} }\sim \frac{|2y_{\mu e \mu e}^{LL}|^2 + |y_{\mu e \mu e}^{LR}|^2}{(2\sqrt{2} G_F)^2} ,
\end{align}
which is at most $10^{-2}$ given the above-derived constraints. 
Because of this, tridents will be at most useful at constraining the weakest linear combination, $|y_{\mu e \mu e}^{LL} - y_{\mu e \mu e}^{RR}|$. For a pure $\nu_\mu$ beam, the $\mu^+ e^-$ appearance would be an unambiguous sign of lepton flavor violation and thus a background-free signature, but realistic neutrino beams will have admixtures of $\nu_e$ and $\bar{\nu}_\mu$ that induce indistinguishable SM processes such as $\nu_e X\to \nu_\mu \mu^+ e^- X$, rendering trident searches for $\Delta L_\mu = 2$ difficult. Still, tridents might eventually become a probe competitive with lepton flavor universality violation.

The $\bar{\mu}\bar{\mu}ee$ operators could also be probed at future lepton colliders via the background-free $e^-e^-\to \mu^- \mu^-$, $\mu^+ e^-\to \mu^- e^+$, or $\mu^+\mu^+\to e^+ e^+$. A setup for the latter two initial states was recently proposed as $\mu$TRISTAN~\cite{Hamada:2022mua},  a  high-energy lepton collider using the ultra-cold antimuon technology developed at J-PARC~\cite{Abe:2019thb} that
could run in the $\mu^+ e^-$ mode with $\sqrt{s} = \unit[346]{GeV}$, and later in the $\mu^+ \mu^+$ mode~\cite{Heusch:1995yw} with $\sqrt{s}=\unit[2]{TeV}$ or even higher.
Judging by the analyses of similar four-lepton operators in Refs.~\cite{Hamada:2022uyn,Fridell:2023gjx}, we can expect $\mu$TRISTAN to probe all $|y_{\mu e \mu e}|$ down to $(\unit[\mathcal{O}(10)]{TeV})^{-2}$, superseding all current limits.
The high centre-of-mass energy might even allow for a direct production of the mediators underlying our $d=6$ operators, see Ref.~\cite{Bossi:2020yne,Yang:2023ojm,Lichtenstein:2023iut,Dev:2023nha} for such studies.
While we have focused on $\mu$TRISTAN here, other collider designs could also provide good reach for $\bar{\mu}\bar{\mu}ee$ operators as long as they collide $\mu^\pm \mu^\pm$, $e^\pm e^\pm$, or $\mu^\pm e^\mp$. $\mu^-$--$e^-$ scattering, e.g.~at MuonE~\cite{Abbiendi:2016xup}, has much weaker sensitivity to our $\Delta L_\mu  = -\Delta L_e = 2$ operators.

\section{\texorpdfstring{$\bar{\tau}\bar{\tau} e e$}{tau-tau-e-e}}
\label{sec:tau-tau-e-e}

Next we consider the $d=6$ operators with $\Delta L_\tau =- \Delta L_e = 2$:
\begin{align}
\mathcal{L}& \supset y^{LL}_{\tau e\tau e}\bar{L}_\tau \gamma^\alpha L_e\, \bar{L}_\tau \gamma_\alpha L_e+ y^{LR}_{\tau e\tau e}\bar{L}_\tau \gamma^\alpha L_e\, \bar{\ell}_\tau \gamma_\alpha \ell_e \nonumber \\
&\quad + y^{RR}_{\tau e\tau e}\bar{\ell}_\tau \gamma^\alpha \ell_e\, \bar{\ell}_\tau \gamma_\alpha \ell_e+ \hc\,.
\end{align}
These operators do not immediately lead to neutrinoless tau decays, nor do we have any tauonium-antitauonium conversion experiments at our disposal. 
For $y^{LL}$ and $y^{LR}$, we can once again calculate the decay rates and electron spectra of $\tau^-\to e^- \bar\nu_\tau\nu_e$ and compare them to experimental data. 
Just like for the muon decay, these operators generate the same electron spectrum as SM tauon decays, so the partial width is simply rescaled compared to the SM:
\begin{align}
    \Gamma_{\tau \to e \nu \nu} = \Gamma_{\tau \to e \nu \nu}^\text{SM} \left(1 + \frac{|2y_{\tau e \tau e}^{LL}|^2 + |y_{\tau e \tau e}^{LR}|^2}{(2\sqrt{2} G_F)^2} \right) .
    \label{eq:GammaTaMuNuNu}
\end{align}
Again, we use lepton-flavor-universality tests to obtain limits. We calculate $\Gamma_{\tau \to e \nu \nu} /\Gamma_{\mu \to e \nu \nu} $ and compare to the experimental values~\cite{HeavyFlavorAveragingGroup:2022wzx}:
\begin{align}
    |2y_{\tau e \tau e}^{LL}|^2 + |y_{\tau e \tau e}^{LR}|^2 =  \unit[(2.43 \pm 3.11)]{TeV^{-4}} \,.
\end{align}
A stronger limit can be achieved using preliminary Belle~II data~\cite{Belle_TAU2023}  (combined with older measurements) for $\Gamma_{\tau \to e \nu \nu}/\Gamma_{\tau \to \mu \nu \nu} $:
\begin{align}
    |2y_{\tau e \tau e}^{LL}|^2 + |y_{\tau e \tau e}^{LR}|^2 = \unit[ (-1.05 \pm 2.90)]{TeV^{-4}} \,.
\end{align}
Using one-sided confidence intervals, we obtain a $95\%$~C.L. limit on $|2y_{\tau e \tau e}^{LL}|$ and $ |y_{\tau e \tau e}^{LR}|$ of $2.2/\unit{TeV^2}$ (Tab.~\ref{tab:limits}). These are viable SMEFT limits that can be improved with future Belle II data.

\begin{figure}[tb]
    \centering
    \includegraphics[width=0.3\textwidth]{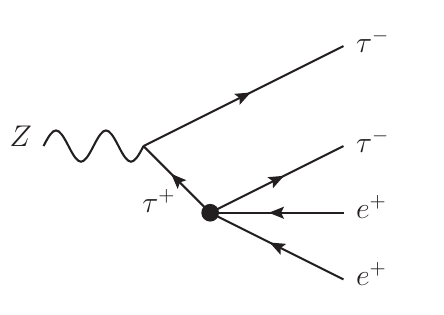}
    \caption{Example Feynman diagram for $Z\to \tau^-\tau^-e^+e^+$ via the $y_{\tau e\tau e}$ operator (filled circle).
    }
    \label{fig:Z_decay}
\end{figure}

$y_{\tau e \tau e}^{RR}$ remains unconstrained here since it does not involve any neutrinos. Closing SM loops to generate neutrinos -- equivalent to mixing the three operators, see Eq.~\eqref{eq:loop} and Fig.~\ref{fig:muemue_mixing} -- requires chirality flips and is thus heavily suppressed.
To avoid such suppressions we can consider  $Z \to \tau\tau \bar{e}\bar{e}$ decays, see Fig.~\ref{fig:Z_decay}. Neglecting lepton masses, the branching ratio for this process is
\begin{align}
    \BR (Z \to \tau^\pm & \tau^\pm e^\mp  e^\mp) \simeq 1.4 \, \frac{M_Z^5}{49152\pi^5}\frac{e^2 s_W^2}{c_W^2 \Gamma_Z} |y^{RR}_{\tau e \tau e}|^2\nonumber \\
    &\simeq  4.18 \times 10^{-11} \left|\frac{y^{RR}_{\tau e \tau e}}{(\unit[0.1]{TeV})^{-2}}\right|^2 \label{eq:Zdecay},
\end{align}
where $M_Z$ ($\Gamma_Z$) is the $Z$ mass (width), and $s_W$ ($c_W$) the sine (cosine) of the weak mixing angle. Currently, we do not have any experimental constraints on this decay channel; demanding the branching ratio to be less than one gives the weak bound $|y^{RR}_{\tau e \tau e}| < 15.5 / \unit{GeV}^{2}$.
The total $Z$ width agrees very well with the SM prediction~\cite{ParticleDataGroup:2022pth}, which can be translated into a $2\sigma$ upper bound of $2\times 10^{-3}$ on any non-SM $Z$ branching ratio.\footnote{The recent $W$-mass measurement by CDF-II~\cite{CDF:2022hxs} is ignored here.} This improves the bound to $|y^{RR}_{\tau e \tau e}| < 0.7 / \unit{GeV}^{2}$ (Tab.~\ref{tab:limits}). 
A dedicated LHC search for this decay could realistically reach branching ratios of order $10^{-5}$, corresponding to a limit $|y^{RR}_{\tau e \tau e}| < 4.89 \times 10^4 / \unit{TeV}^{2}$.
$Z$ decays could conceivably be measured more extensively in the future at a so-called $Z$ factory~\cite{Bernardi:2022hny}, producing trillions of $Z$ bosons. Assuming an optimistic reach of  $10^{-12}$ for the above branching ratios would probe  $|y^{RR}_{\tau e \tau e}| < 15.5 / \unit{TeV}^{2}$, just barely above the electroweak scale.
This is likely our best shot at providing EFT limits on $y_{\tau e \tau e}^{RR}$. Better limits on explicit UV completions of this operator are of course possible and probe complementary parameter space, see Ref.~\cite{Altmannshofer:2016brv,Altmannshofer:2022fvz,Bigaran:2022giz}.
Notice that four-lepton $Z$ decays are also useful for many other SMEFT coefficients~\cite{Boughezal:2020klp}.

\begin{figure}[tb]
    \centering
    \includegraphics[width=0.4\textwidth]{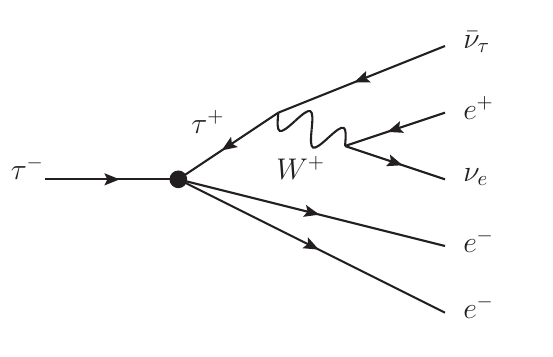}
    \caption{Example Feynman diagram for $\tau^-\to e^-e^- e^+ \nu_e \bar{\nu}_\tau$ via the $y_{\tau e\tau e}$ operator (filled circle).
    }
    \label{fig:five_body_tau_decay}
\end{figure}

Yet another probe of the $RR$ coupling can be found in tauon decays involving an off-shell tauon in the final state, see Fig.~\ref{fig:five_body_tau_decay}. These are additionally suppressed by $G_F$ and phase space, but can be competitive due to the large number of collected tauon decays compared to $Z$. For massless final states we find
\begin{align}
    \BR (\tau^-&\to e^- e^- \ell^+ \nu_\ell \bar{\nu}_\tau) \simeq  1.1\,\frac{m_\tau^9 G_F^2}{4718592\pi^7 \Gamma_\tau} |y^{RR}_{\tau e \tau e}|^2\nonumber \\
    &\simeq  8.2 \times 10^{-15} \left|\frac{y^{RR}_{\tau e \tau e}}{(\unit[0.1]{TeV})^{-2}}\right|^2 \label{eq:off-shell_tau_decay},
\end{align}
where $\ell$ is an electron or muon;  $\tau^- \to e^- e^- \pi^+ \bar{\nu}_\tau$ has a similar rate.
Probing scales above the electroweak scale is clearly out of the question even at Belle II, but at least we can confirm the current $Z$-decay limit from above. For this we notice that CLEO has long ago observed the SM decay $\tau^- \to e^- e^- e^+ \nu_\tau\bar{\nu}_e$~\cite{CLEO:1995azm} with branching ratio $(2.8\pm 1.5)\times 10^{-5}$~\cite{CLEO:1995azm}, compatible with the SM prediction $4.2\times 10^{-5}$~\cite{Dicus:1994dt,CLEO:1995azm,Flores-Tlalpa:2015vga}. 
This yields a $95\%$ C.L.~limit of $4.4/\unit{GeV}^2$ on $|y^{RR}_{\tau e \tau e}|$, slightly worse than the $Z$-decay limit but with entirely different assumptions.
Belle II can significantly improve this limit.

\section{\texorpdfstring{$\bar{\tau}\bar{\tau} \mu \mu$}{tau-tau-mu-mu}}
\label{sec:tau-tau-mu-mu}

In this section we consider the $d=6$ operators with $\Delta L_\tau =- \Delta L_\mu = 2$, 
\begin{align}
\mathcal{L} & \supset y^{LL}_{\tau \mu\tau \mu}\bar{L}_\tau \gamma^\alpha L_\mu\, \bar{L}_\tau \gamma_\alpha L_\mu+ y^{LR}_{\tau \mu\tau \mu}\bar{L}_\tau \gamma^\alpha L_\mu\, \bar{\ell}_\tau \gamma_\alpha \ell_\mu \nonumber \\
&\quad + y^{RR}_{\tau \mu\tau \mu}\bar{\ell}_\tau \gamma^\alpha \ell_\mu\, \bar{\ell}_\tau \gamma_\alpha \ell_\mu+ \hc\,,
\end{align}
which have a similar phenomenology as the $\bar{\tau}\bar{\tau} e e$ operators from the previous section.
$y_{\tau \mu \tau \mu}^{LL}$ and $ y_{\tau \mu \tau \mu}^{LR}$ give rise to $\tau^-\to \mu^- \nu_\mu\bar\nu_\tau $, once again just rescaling the SM decay rate. We compute $\Gamma_{\tau \to \mu \nu \nu} /\Gamma_{\mu \to e \nu \nu} $ and compare it to the experimental results,  obtaining
\begin{align}
    |2y_{\tau \mu \tau \mu}^{LL}|^2 + |y_{\tau \mu \tau \mu}^{LR}|^2 = \unit[\left(6.14 \pm 3.17\right) ]{TeV^{-4}} \,.
\end{align}
A better limit can be achieved using preliminary Belle~II data (combined with older measurements)~\cite{Belle_TAU2023}  for $\Gamma_{\tau \to \mu \nu \nu} /\Gamma_{\tau \to e \nu \nu} $:
\begin{align}
    |2y_{\tau \mu \tau \mu}^{LL}|^2 + |y_{\tau \mu \tau \mu}^{LR}|^2 = \left(1.05 \pm 2.91\right) \unit{TeV^{-4}} \,.
\end{align}
Using one-sided confidence intervals, one can get $95\%$~C.L. limits on $|2y_{\tau \mu \tau \mu}^{LL}|$ and $ |y_{\tau \mu \tau \mu}^{LR}|$ of $2.5/\unit{TeV^2}$ (Tab.~\ref{tab:limits}).

Just like in the previous section, the last Wilson coefficient $y_{\tau \mu \tau \mu}^{RR}$  can be constrained using $Z \to \tau\tau \bar{\mu}\bar{\mu}$ decays (Tab.~\ref{tab:limits}). Since we are considering tauons, muons, and electrons to be massless we obtain the same branching ratio as in Eq.~\eqref{eq:Zdecay} and the same current and future limits.
The $\bar{\tau}\bar{\tau}\mu\mu$ operators could also be probed at $\mu$TRISTAN via $\mu^+\mu^+\to\tau^+\tau^+$, with sensitivity to all $|y_{
\tau\mu \tau\mu}|$ down to $(\unit[\mathcal{O}(10)]{TeV})^{-2}$.

Five-body tauon decays $\tau^- \to \mu^- \mu^- \ell^+ \nu_\ell \bar{\nu}_\tau$ can be obtained from Eq.~\eqref{eq:off-shell_tau_decay} with $e\to \mu$, but in this case there is no experimental data to compare to.

\section{\texorpdfstring{$\bar{\tau}\bar{\tau} e \mu$}{tau-tau-e-mu}}

The last three sections deal with lepton-triality-allowed operators, first the ones with $\Delta L_\tau= -2 \Delta L_\mu =-2 \Delta L_e = 2$:
\begin{align}
\mathcal{L}& \supset y^{LL}_{e\tau \mu \tau}\bar{L}_e \gamma^\alpha L_{\tau}\, \bar{L}_\mu \gamma_\alpha L_{\tau}\nonumber \\
&\quad + y^{LR}_{e\tau \mu \tau}\bar{L}_e \gamma^\alpha L_{\tau}\, \bar{\ell}_\mu \gamma_\alpha \ell_{\tau}
+ y^{LR}_{\mu \tau e \tau}\bar{L}_\mu \gamma^\alpha L_{\tau}\, \bar{\ell}_e \gamma_\alpha \ell_{\tau} \nonumber \\
&\quad + y^{RR}_{e\tau \mu \tau}\bar{\ell}_e \gamma^\alpha \ell_{\tau}\, \bar{\ell}_\mu \gamma_\alpha \ell_{\tau} + \hc
\end{align}
These terms lead to two different tau decay channels: $\tau^-\to e^- \nu_\mu\bar\nu_\tau$ from $y_{e \tau \mu \tau}^{LL}$ and $ y_{\mu \tau e \tau}^{LR}$,  and $\tau^-\to \mu^- \nu_e\bar\nu_\tau$ from $y_{e \tau \mu \tau}^{LL}$ and $ y_{e \tau \mu \tau}^{LR}$:
\begin{align}
    \Gamma_{\tau \to e \nu\nu} &= \Gamma_{\tau \to e \nu\nu}^\text{SM} \left(1 + \frac{ |y_{e \tau \mu \tau}^{LL}|^2 + |y_{\mu \tau e \tau}^{LR}|^2}{(2\sqrt{2} G_F)^2} \right) \label{eq:tauToE5},\\
    \Gamma_{\tau \to \mu \nu\nu} &= \Gamma_{\tau \to \mu \nu\nu}^\text{SM} \left(1 + \frac{|y_{e \tau \mu \tau}^{LL}|^2 + |y_{e \tau \mu \tau}^{LR}|^2}{(2\sqrt{2} G_F)^2} \right) \label{eq:tauToMu5}.
\end{align}
We implement the by now familiar lepton-flavor-universality test and compare the above rates to $\Gamma (\mu \to e\nu\nu)$ to obtain:
\begin{align}
    |y_{e \tau \mu \tau}^{LL}|^2 + |y_{\mu \tau e \tau}^{LR}|^2 = (2.43 \pm 3.11) \unit{TeV}^{-4} \, \label{eq:eTauMuTau}, \\
    |y_{e \tau \mu \tau}^{LL}|^2 + |y_{e \tau \mu \tau}^{LR}|^2 = (6.14 \pm 3.17) \unit{TeV}^{-4} \,.
\end{align}
Using one-sided confidence intervals, one can get $95\%$~C.L. limits of $7.9/\unit{TeV^4}$ (Tab.~\ref{tab:limits}) and $11/\unit{TeV^4}$ on $|y_{e \tau \mu \tau}^{LL}|^2 + |y_{\mu \tau e \tau}^{LR}|^2$ and $|y_{e \tau \mu \tau}^{LL}|^2 + |y_{e \tau \mu \tau}^{LR}|^2$ respectively.

Finally, we compare Eq.~\eqref{eq:tauToE5} to Eq.~\eqref{eq:tauToMu5}:
\begin{align}
    \frac{\Gamma_{\tau \to e \nu\nu}}{\Gamma_{\tau \to \mu \nu\nu}} \simeq \frac{\Gamma_{\tau \to e \nu\nu}^\text{SM}}{\Gamma_{\tau \to \mu \nu\nu}^\text{SM}} \left( 1 + \frac{|y_{\mu \tau e \tau}^{LR}|^2 - |y_{e \tau \mu \tau}^{LR}|^2}{(2\sqrt{2} G_F)^2} \right).
\end{align}
By comparing to the experimental data from Eq.~\eqref{eq:belleII} we get a limit on the following linear combination:
\begin{align}
    |y_{\mu \tau e \tau}^{LR}|^2 - |y_{e \tau \mu \tau}^{LR}|^2 = \unit[(-1.05 \pm 2.90)]{TeV^{-4}}.
    \label{eq:eTauMuTauMinus}
\end{align}
This allows us to improve the limit on $|y_{e \tau \mu \tau}^{LR}|$. Combining equations \eqref{eq:eTauMuTau} and \eqref{eq:eTauMuTauMinus} we have:
\begin{align}
\begin{split}
    |y_{e \tau \mu \tau}^{LR}|^2 &=  |y_{\mu \tau e \tau}^{LR}|^2 - \left( |y_{\mu \tau e \tau}^{LR}|^2 - |y_{e \tau \mu \tau}^{LR}|^2 \right)\\
    & \leq  (3.48 \pm 4.25)\unit{TeV}^{-4} \,.
\end{split}
\end{align}
Again, utilizing one-sided confidence intervals we recover a slightly improved limit on $|y_{e \tau \mu \tau}^{LR}|$ of $3.3/\unit{TeV}^2$ (Tab.~\ref{tab:limits}).

Three of the four Wilson coefficients are thus constrained from universality ratios. 
Similarly to sections \ref{sec:tau-tau-e-e} and \ref{sec:tau-tau-mu-mu} one can obtain limits on $|y_{e \tau \mu \tau}^{RR}|$ from $Z$ decays. The prefactor of the branching ratio is a factor of 2 smaller here since two final state particles are not the same anymore:
\begin{align}
\BR (Z \to \tau^\pm\tau^\pm & e^\mp \mu^\mp) \simeq 2.09 \times 10^{-11} \left|\frac{y^{RR}_{ e \tau \mu \tau}}{(\unit[0.1]{TeV})^{-2}}\right|^2.
\label{eq:brZTauTauEMu}
\end{align}
This change makes the limit on $|y_{e \tau \mu \tau}^{RR}|$ a factor of $\sqrt{2}$ weaker compared to $|y^{RR}_{\tau e \tau e}|$ and $|y^{RR}_{\tau \mu \tau \mu}|$ (Tab.~\ref{tab:limits}).

Five-body tauon decays $\tau^- \to e^- \mu^- \ell^+ \nu_\ell \bar{\nu}_\tau$ also have a factor-2 smaller branching ratio than Eq.~\eqref{eq:off-shell_tau_decay}:
\begin{align}
    \BR (\tau^-&\to e^- \mu^- \ell^+ \nu_\ell \bar{\nu}_\tau) \simeq  4.1 \times 10^{-15} \left|\frac{y^{RR}_{e\tau \mu \tau}}{(\unit[0.1]{TeV})^{-2}}\right|^2 .
    \label{eq:brTauEMuLepNuNu}
\end{align}
CLEO provides a $90\%$~C.L.~limit on the SM-allowed decay
$\BR(\tau^- \to \mu^- e^- e^+ \nu_\tau\bar{\nu}_\mu) < 3.2\times 10^{-5}$~\cite{CLEO:1995azm} which translates to a bound on $|y^{RR}_{e\tau \mu \tau}| < 5.4/\unit{GeV}^{2}$.

\section{\texorpdfstring{$\bar{\tau}\bar{\mu} e e$}{tau-mu-e-e}}

Now we investigate $\Delta L_e= -2 \Delta L_\tau =-2 \Delta L_\mu = 2$:
\begin{align}
\mathcal{L}& \supset y^{LL}_{\mu e\tau e}\bar{L}_\mu \gamma^\alpha L_e\, \bar{L}_\tau \gamma_\alpha L_e \nonumber \\
&\quad + y^{LR}_{\mu e\tau e}\bar{L}_\mu \gamma^\alpha L_e\, \bar{\ell}_\tau \gamma_\alpha \ell_e
+ y^{LR}_{\tau e\mu e}\bar{L}_\tau \gamma^\alpha L_e\, \bar{\ell}_\mu \gamma_\alpha \ell_e \nonumber \\
&\quad + y^{RR}_{\mu e\tau e}\bar{\ell}_\mu \gamma^\alpha \ell_e\, \bar{\ell}_\tau \gamma_\alpha \ell_e+ \hc\,,
\end{align}
all of which give rise to the clean lepton-flavor-violating decay $\tau^+ \to e^+e^+ \mu^-$ with rate
\begin{align}
    \Gamma \simeq \frac{m_\tau^5\left(|y^{LL}_{\mu e\tau e}|^2 + |y^{LR}_{\mu e\tau e}|^2 + |y^{LR}_{\tau e\mu e}|^2 + |y^{RR}_{\tau e\mu e}|^2\right)}{1536 \pi ^3}\,,
\end{align}
assuming vanishing electron and muon mass and hence no interference terms. 
Dedicated searches for this decay mode at Belle~\cite{Hayasaka:2010np}
 yield the strong limits  $|y^{LL}_{\mu e\tau e}|, \dots ,|y^{RR}_{\tau e\mu e}| < 0.0096/\text{TeV}^2$ (Tab.~\ref{tab:limits}).
Belle~II is expected to reach $\BR(\tau^- \to \mu^+ e^- e^-) < 2.3 \times 10^{-10}$  with $\unit[50]{ab^{-1}}$~\cite{Belle-II:2022cgf,Banerjee:2022vdd}, which  can probe the Wilson coefficients down to $ (\unit[29.0]{TeV})^{-2}$.

The lepton-triality-allowed operators $\bar{\tau}\bar{\mu} e e$ have also recently been investigated in Ref.~\cite{Lichtenstein:2023iut}, where it was shown that $\mu^+ e^-\to e^+\tau^-$ at $\mu$TRISTAN would likely provide weaker constraints on $|y_{\mu e\tau e,\tau e\mu e}|$ than Belle II through $\tau^+ \to e^+e^+ \mu^-$.

\section{\texorpdfstring{$\bar{\mu}\bar{\mu} e \tau$}{mu-mu-e-tau}}

Finally, we look at terms with $\Delta L_\mu= -2 \Delta L_e =2 \Delta L_\tau = 2$:
\begin{align}
\mathcal{L}& \supset y^{LL}_{e\mu \tau \mu}\bar{L}_e \gamma^\alpha L_{\mu}\, \bar{L}_\tau \gamma_\alpha L_{\mu} \nonumber \\
&\quad + y^{LR}_{e\mu \tau \mu}\bar{L}_e \gamma^\alpha L_{\mu}\, \bar{\ell}_\tau \gamma_\alpha \ell_{\mu}
+ y^{LR}_{\tau \mu e \mu}\bar{L}_\tau \gamma^\alpha L_{\mu}\, \bar{\ell}_e \gamma_\alpha \ell_{\mu} \nonumber \\
&\quad + y^{RR}_{e\mu \tau \mu}\bar{\ell}_e \gamma^\alpha \ell_{\mu}\, \bar{\ell}_\tau \gamma_\alpha \ell_{\mu} + \hc\,,
\end{align}
which induce $\tau^+ \to \mu^+\mu^+ e^-$
with rate
\begin{align}
\hspace{-1ex}    \Gamma \simeq \frac{m_\tau^5\left(|y^{LL}_{e \mu \tau \mu}|^2 + |y^{LR}_{e \mu \tau \mu}|^2 + |y^{LR}_{\tau \mu e  \mu}|^2 + |y^{RR}_{e \mu \tau \mu}|^2\right)}{1536 \pi ^3}\,.
\end{align}
Comparison with Belle data~\cite{Hayasaka:2010np} puts the following constraints on all four Wilson coefficients $|y^{LL}_{e \mu \tau \mu}|, \dots ,|y^{RR}_{e \mu \tau \mu}| < 0.013/\text{TeV}^2$ (Tab.~\ref{tab:limits}).
Belle II should reach  $\BR(\tau^- \to \mu^- \mu^-  e^+) < 2.6 \times 10^{-10}$~\cite{Belle-II:2022cgf,Banerjee:2022vdd}, which translates to $|y| < (\unit[27.9]{TeV})^{-2}$.

Just like in the previous section, these lepton-triality-allowed operators $\bar{\mu}\bar{\mu} e \tau$ have recently been investigated in Ref.~\cite{Lichtenstein:2023iut}. Here, $\mu^+ \mu^+\to e^+\tau^+$ at $\mu$TRISTAN can be competitive with Belle II through $\tau^+ \to \mu^+\mu^+ e^-$ and reach $|y| < (\unit[\mathcal{O}(10)]{TeV})^{-2}$.

\vspace{1ex}

\section{Conclusions}
\label{sec:conclusions}

Tests of the SM's predicted lepton flavor conservation are among the best probes of new physics, notably in neutrinoless decays of muons and tauons.
Not all lepton flavor violation comes with such clean signatures though: effective operators with $\Delta L_\alpha =2$ are much harder to probe, even though they already arise at mass dimension $d=6$ in the SMEFT and, of course, violate lepton flavor. 8 of these 21 operators give rise to the clean neutrinoless decays $\tau^-\to e^- e^-\mu^+, \mu^-\mu^- e^+$, while two more induce muonium--antimuonium conversion. The remaining 11 operators are rarely discussed, presumably because they are harder to detect.  We have shown that 8 of them can be tested through lepton-flavor-universality violations, i.e.~by comparing leptonic decay rates involving neutrinos, see Tab.~\ref{tab:limits}. This leaves only 3 operators that are currently unconstrained, or at least with such weak constraints that the use of effective field theory is questionable; these require future colliders for unambiguous tests, either in the form of a $Z$ factory or like-sign electron or muon colliders. Most of the other operators will be tested more thoroughly at Belle II via searches for $\tau^-\to e^- e^-\mu^+, \mu^-\mu^- e^+$ as well as improved measurements of $\tau\to e\nu\nu$ and $\tau\to\mu\nu\nu $ that feed into lepton-flavor-universality tests.
Upcoming muonium experiments such as MACE take care of two $\bar{\mu}\bar{\mu}ee$ operators.
The present study serves as a reminder that lepton flavor violation could still be hidden at low scales in comparably murky observables.


\section*{Acknowledgements}
This work was supported in part by the National Science Foundation under Grant PHY-2210428.


\bibliographystyle{utcaps_mod}
\bibliography{BIB.bib}

\providecommand{\href}[2]{#2}\begingroup\raggedright\begin{thebibliography}{10}

\bibitem{Calibbi:2017uvl}
L.~Calibbi and G.~Signorelli, ``{Charged Lepton Flavour Violation: An
  Experimental and Theoretical Introduction},''
  \href{http://dx.doi.org/10.1393/ncr/i2018-10144-0}{{\em Riv. Nuovo Cim.}
  {\bfseries 41} no.~2, (2018) 71--174},
  \href{http://arxiv.org/abs/1709.00294}{[{\ttfamily 1709.00294}]}.

\bibitem{Davidson:2022jai}
S.~Davidson, B.~Echenard, R.~H. Bernstein, J.~Heeck, and D.~G. Hitlin,
  ``{Charged Lepton Flavor Violation},''
  \href{http://arxiv.org/abs/2209.00142}{[{\ttfamily 2209.00142}]}.

\bibitem{Heeck:2016xwg}
J.~Heeck, ``{Interpretation of Lepton Flavor Violation},''
  \href{http://dx.doi.org/10.1103/PhysRevD.95.015022}{{\em Phys. Rev. D}
  {\bfseries 95} (2017) 015022},
  \href{http://arxiv.org/abs/1610.07623}{[{\ttfamily 1610.07623}]}.

\bibitem{Isidori:2023pyp}
G.~Isidori, F.~Wilsch, and D.~Wyler, ``{The standard model effective field
  theory at work},'' \href{http://dx.doi.org/10.1103/RevModPhys.96.015006}{{\em
  Rev. Mod. Phys.} {\bfseries 96} no.~1, (2024) 015006},
  \href{http://arxiv.org/abs/2303.16922}{[{\ttfamily 2303.16922}]}.

\bibitem{Weinberg:1979sa}
S.~Weinberg, ``{Baryon and Lepton Nonconserving Processes},''
  \href{http://dx.doi.org/10.1103/PhysRevLett.43.1566}{{\em Phys. Rev. Lett.}
  {\bfseries 43} (1979) 1566--1570}.

\bibitem{Weinberg:1980bf}
S.~Weinberg, ``{Varieties of Baryon and Lepton Nonconservation},''
  \href{http://dx.doi.org/10.1103/PhysRevD.22.1694}{{\em Phys. Rev. D}
  {\bfseries 22} (1980) 1694}.

\bibitem{Buchmuller:1985jz}
W.~Buchmuller and D.~Wyler, ``{Effective Lagrangian Analysis of New
  Interactions and Flavor Conservation},''
  \href{http://dx.doi.org/10.1016/0550-3213(86)90262-2}{{\em Nucl. Phys. B}
  {\bfseries 268} (1986) 621--653}.

\bibitem{Grzadkowski:2010es}
B.~Grzadkowski, M.~Iskrzynski, M.~Misiak, and J.~Rosiek, ``{Dimension-Six Terms
  in the Standard Model Lagrangian},''
  \href{http://dx.doi.org/10.1007/JHEP10(2010)085}{{\em JHEP} {\bfseries 10}
  (2010) 085}, \href{http://arxiv.org/abs/1008.4884}{[{\ttfamily 1008.4884}]}.

\bibitem{Rodejohann:2011mu}
W.~Rodejohann, ``{Neutrino-less Double Beta Decay and Particle Physics},''
  \href{http://dx.doi.org/10.1142/S0218301311020186}{{\em Int. J. Mod. Phys. E}
  {\bfseries 20} (2011) 1833--1930},
  \href{http://arxiv.org/abs/1106.1334}{[{\ttfamily 1106.1334}]}.

\bibitem{Cuypers:1996ia}
F.~Cuypers and S.~Davidson, ``{Bileptons: Present limits and future
  prospects},'' \href{http://dx.doi.org/10.1007/s100520050157}{{\em Eur. Phys.
  J. C} {\bfseries 2} (1998) 503--528},
  \href{http://arxiv.org/abs/hep-ph/9609487}{[{\ttfamily hep-ph/9609487}]}.

\bibitem{Altmannshofer:2016brv}
W.~Altmannshofer, C.-Y. Chen, P.~S. Bhupal~Dev, and A.~Soni, ``{Lepton flavor
  violating Z' explanation of the muon anomalous magnetic moment},''
  \href{http://dx.doi.org/10.1016/j.physletb.2016.09.046}{{\em Phys. Lett. B}
  {\bfseries 762} (2016) 389--398},
  \href{http://arxiv.org/abs/1607.06832}{[{\ttfamily 1607.06832}]}.

\bibitem{Fukuyama:2021iyw}
T.~Fukuyama, Y.~Mimura, and Y.~Uesaka, ``{Models of the muonium to antimuonium
  transition},'' \href{http://dx.doi.org/10.1103/PhysRevD.105.015026}{{\em
  Phys. Rev. D} {\bfseries 105} (2022) 015026},
  \href{http://arxiv.org/abs/2108.10736}{[{\ttfamily 2108.10736}]}.

\bibitem{Goudelis:2023yni}
A.~Goudelis, J.~Kriewald, E.~Pinsard, and A.~M. Teixeira, ``{cLFV leptophilic
  $Z^\prime$ as a dark matter portal: prospects for colliders},''
  \href{http://arxiv.org/abs/2312.14103}{[{\ttfamily 2312.14103}]}.

\bibitem{Ma:2010gs}
E.~Ma, ``{Quark and Lepton Flavor Triality},''
  \href{http://dx.doi.org/10.1103/PhysRevD.82.037301}{{\em Phys. Rev. D}
  {\bfseries 82} (2010) 037301},
  \href{http://arxiv.org/abs/1006.3524}{[{\ttfamily 1006.3524}]}.

\bibitem{Cao:2011df}
Q.-H. Cao, A.~Damanik, E.~Ma, and D.~Wegman, ``{Probing Lepton Flavor Triality
  with Higgs Boson Decay},''
  \href{http://dx.doi.org/10.1103/PhysRevD.83.093012}{{\em Phys. Rev. D}
  {\bfseries 83} (2011) 093012},
  \href{http://arxiv.org/abs/1103.0008}{[{\ttfamily 1103.0008}]}.

\bibitem{Bigaran:2022giz}
I.~Bigaran, X.-G. He, M.~A. Schmidt, G.~Valencia, and R.~Volkas,
  ``{Lepton-flavor-violating tau decays from triality},''
  \href{http://dx.doi.org/10.1103/PhysRevD.107.055001}{{\em Phys. Rev. D}
  {\bfseries 107} (2023) 055001},
  \href{http://arxiv.org/abs/2212.09760}{[{\ttfamily 2212.09760}]}.

\bibitem{Altarelli:2005yx}
G.~Altarelli and F.~Feruglio, ``{Tri-bimaximal neutrino mixing, $A(4)$ and the
  modular symmetry},''
  \href{http://dx.doi.org/10.1016/j.nuclphysb.2006.02.015}{{\em Nucl. Phys. B}
  {\bfseries 741} (2006) 215--235},
  \href{http://arxiv.org/abs/hep-ph/0512103}{[{\ttfamily hep-ph/0512103}]}.

\bibitem{Ishimori:2010au}
H.~Ishimori, T.~Kobayashi, H.~Ohki, Y.~Shimizu, H.~Okada, and M.~Tanimoto,
  ``{Non-Abelian Discrete Symmetries in Particle Physics},''
  \href{http://dx.doi.org/10.1143/PTPS.183.1}{{\em Prog. Theor. Phys. Suppl.}
  {\bfseries 183} (2010) 1--163},
  \href{http://arxiv.org/abs/1003.3552}{[{\ttfamily 1003.3552}]}.

\bibitem{Holthausen:2012wz}
M.~Holthausen, M.~Lindner, and M.~A. Schmidt, ``{Lepton flavor at the
  electroweak scale: A complete $A_{4}$ model},''
  \href{http://dx.doi.org/10.1103/PhysRevD.87.033006}{{\em Phys. Rev. D}
  {\bfseries 87} (2013) 033006},
  \href{http://arxiv.org/abs/1211.5143}{[{\ttfamily 1211.5143}]}.

\bibitem{Willmann:1998gd}
L.~Willmann {\em et~al.}, ``{New bounds from searching for muonium to
  anti-muonium conversion},''
  \href{http://dx.doi.org/10.1103/PhysRevLett.82.49}{{\em Phys. Rev. Lett.}
  {\bfseries 82} (1999) 49--52},
  \href{http://arxiv.org/abs/hep-ex/9807011}{[{\ttfamily hep-ex/9807011}]}.

\bibitem{HeavyFlavorAveragingGroup:2022wzx}
{\bfseries Heavy Flavor Averaging Group, HFLAV} Collaboration, Y.~S. Amhis {\em
  et~al.}, ``{Averages of b-hadron, c-hadron, and \ensuremath{\tau}-lepton
  properties as of 2021},''
  \href{http://dx.doi.org/10.1103/PhysRevD.107.052008}{{\em Phys. Rev. D}
  {\bfseries 107} (2023) 052008},
  \href{http://arxiv.org/abs/2206.07501}{[{\ttfamily 2206.07501}]}.

\bibitem{Belle_TAU2023}
{\bfseries Belle II} Collaboration, P.~Feichtinger, ``{Measurements of Michel
  parameters and tests of lepton universality in $\tau$ decays at Belle and
  Belle II},'' {\em Talk at the TAU2023 conference in Louisville, Kentucky,
  \url{https://indico.cern.ch/event/1303630/contributions/5571692/attachments/2765068/4816167/tau2023_belle12_michel_LFU.pdf}}
  (2023) .

\bibitem{Hayasaka:2010np}
K.~Hayasaka {\em et~al.}, ``{Search for Lepton Flavor Violating Tau Decays into
  Three Leptons with 719 Million Produced Tau+Tau- Pairs},''
  \href{http://dx.doi.org/10.1016/j.physletb.2010.03.037}{{\em Phys. Lett. B}
  {\bfseries 687} (2010) 139--143},
  \href{http://arxiv.org/abs/1001.3221}{[{\ttfamily 1001.3221}]}.

\bibitem{Conlin:2020veq}
R.~Conlin and A.~A. Petrov, ``{Muonium-antimuonium oscillations in effective
  field theory},'' \href{http://dx.doi.org/10.1103/PhysRevD.102.095001}{{\em
  Phys. Rev. D} {\bfseries 102} (2020) 095001},
  \href{http://arxiv.org/abs/2005.10276}{[{\ttfamily 2005.10276}]}.

\bibitem{Petrov:2022wau}
A.~A. Petrov, R.~Conlin, and C.~Grant, ``{Studying $\Delta L=2$ Lepton Flavor
  Violation with Muons},''
  \href{http://dx.doi.org/10.3390/universe8030169}{{\em Universe} {\bfseries 8}
  no.~3, (2022) 169}, \href{http://arxiv.org/abs/2203.04161}{[{\ttfamily
  2203.04161}]}.

\bibitem{Fukuyama:2022fwi}
T.~Fukuyama, Y.~Mimura, and Y.~Uesaka, ``{Transverse positron polarization in
  the polarized $\mu^+$ decay related with the muonium-to-antimuonium
  transition},'' \href{http://dx.doi.org/10.1103/PhysRevD.105.075024}{{\em
  Phys. Rev. D} {\bfseries 105} (2022) 075024},
  \href{http://arxiv.org/abs/2201.06279}{[{\ttfamily 2201.06279}]}.

\bibitem{Fukuyama:2023drl}
T.~Fukuyama, Y.~Mimura, and Y.~Uesaka, ``{Insights from the magnetic field
  dependence of the muonium-to-antimuonium transition},''
  \href{http://dx.doi.org/10.1103/PhysRevD.108.095029}{{\em Phys. Rev. D}
  {\bfseries 108} (2023) 095029},
  \href{http://arxiv.org/abs/2309.02060}{[{\ttfamily 2309.02060}]}.

\bibitem{Bai:2022sxq}
A.-Y. Bai {\em et~al.}, ``{Snowmass2021 Whitepaper: Muonium to antimuonium
  conversion},'' in {\em {Snowmass 2021}}.
\newblock 3, 2022.
\newblock \href{http://arxiv.org/abs/2203.11406}{[{\ttfamily 2203.11406}]}.

\bibitem{Kawamura:2021lqk}
N.~Kawamura, R.~Kitamura, H.~Yasuda, M.~Otani, Y.~Nakazawa, H.~Iinuma, and
  T.~Mibe, ``{A new approach for Mu \ensuremath{-} \(\overline{\text{Mu}}\)
  Conversion Search},'' \href{http://dx.doi.org/10.7566/JPSCP.33.011120}{{\em
  JPS Conf. Proc.} {\bfseries 33} (2021) 011120}.

\bibitem{Jenkins:2013wua}
E.~E. Jenkins, A.~V. Manohar, and M.~Trott, ``{Renormalization Group Evolution
  of the Standard Model Dimension Six Operators II: Yukawa Dependence},''
  \href{http://dx.doi.org/10.1007/JHEP01(2014)035}{{\em JHEP} {\bfseries 01}
  (2014) 035}, \href{http://arxiv.org/abs/1310.4838}{[{\ttfamily 1310.4838}]}.

\bibitem{Fujii:1993su}
H.~Fujii, Y.~Mimura, K.~Sasaki, and T.~Sasaki, ``{Muonium, hyperfine structure
  and the decay $\mu^+\to e^+ + \bar{\nu}_e + \nu_\mu$ in models with dilepton
  gauge baryons},'' \href{http://dx.doi.org/10.1103/PhysRevD.49.559}{{\em Phys.
  Rev. D} {\bfseries 49} (1994) 559--562},
  \href{http://arxiv.org/abs/hep-ph/9309287}{[{\ttfamily hep-ph/9309287}]}.

\bibitem{CDF:2022hxs}
{\bfseries CDF} Collaboration, T.~Aaltonen {\em et~al.}, ``{High-precision
  measurement of the W boson mass with the CDF II detector},''
  \href{http://dx.doi.org/10.1126/science.abk1781}{{\em Science} {\bfseries
  376} no.~6589, (2022) 170--176}.

\bibitem{Magill:2016hgc}
G.~Magill and R.~Plestid, ``{Neutrino Trident Production at the Intensity
  Frontier},'' \href{http://dx.doi.org/10.1103/PhysRevD.95.073004}{{\em Phys.
  Rev. D} {\bfseries 95} (2017) 073004},
  \href{http://arxiv.org/abs/1612.05642}{[{\ttfamily 1612.05642}]}.

\bibitem{Magill:2017mps}
G.~Magill and R.~Plestid, ``{Probing new charged scalars with neutrino trident
  production},'' \href{http://dx.doi.org/10.1103/PhysRevD.97.055003}{{\em Phys.
  Rev. D} {\bfseries 97} (2018) 055003},
  \href{http://arxiv.org/abs/1710.08431}{[{\ttfamily 1710.08431}]}.

\bibitem{Ballett:2018uuc}
P.~Ballett, M.~Hostert, S.~Pascoli, Y.~F. Perez-Gonzalez, Z.~Tabrizi, and
  R.~Zukanovich~Funchal, ``{Neutrino Trident Scattering at Near Detectors},''
  \href{http://dx.doi.org/10.1007/JHEP01(2019)119}{{\em JHEP} {\bfseries 01}
  (2019) 119}, \href{http://arxiv.org/abs/1807.10973}{[{\ttfamily
  1807.10973}]}.

\bibitem{Altmannshofer:2019zhy}
W.~Altmannshofer, S.~Gori, J.~Mart\'\i{}n-Albo, A.~Sousa, and M.~Wallbank,
  ``{Neutrino Tridents at DUNE},''
  \href{http://dx.doi.org/10.1103/PhysRevD.100.115029}{{\em Phys. Rev. D}
  {\bfseries 100} (2019) 115029},
  \href{http://arxiv.org/abs/1902.06765}{[{\ttfamily 1902.06765}]}.

\bibitem{Hamada:2022mua}
Y.~Hamada, R.~Kitano, R.~Matsudo, H.~Takaura, and M.~Yoshida,
  ``{$\mu$TRISTAN},'' \href{http://dx.doi.org/10.1093/ptep/ptac059}{{\em PTEP}
  {\bfseries 2022} no.~5, (2022) 053B02},
  \href{http://arxiv.org/abs/2201.06664}{[{\ttfamily 2201.06664}]}.

\bibitem{Abe:2019thb}
M.~Abe {\em et~al.}, ``{A New Approach for Measuring the Muon Anomalous
  Magnetic Moment and Electric Dipole Moment},''
  \href{http://dx.doi.org/10.1093/ptep/ptz030}{{\em PTEP} {\bfseries 2019}
  no.~5, (2019) 053C02}, \href{http://arxiv.org/abs/1901.03047}{[{\ttfamily
  1901.03047}]}.

\bibitem{Heusch:1995yw}
C.~A. Heusch and F.~Cuypers, ``{Physics with like-sign muon beams in a TeV muon
  collider},'' \href{http://dx.doi.org/10.1063/1.49345}{{\em AIP Conf. Proc.}
  {\bfseries 352} (1996) 219--231},
  \href{http://arxiv.org/abs/hep-ph/9508230}{[{\ttfamily hep-ph/9508230}]}.

\bibitem{Hamada:2022uyn}
Y.~Hamada, R.~Kitano, R.~Matsudo, and H.~Takaura, ``{Precision $\mu^+\mu^+$ and
  $\mu^+ e^-$ elastic scatterings},''
  \href{http://dx.doi.org/10.1093/ptep/ptac174}{{\em PTEP} {\bfseries 2023}
  (2023) 013B07}, \href{http://arxiv.org/abs/2210.11083}{[{\ttfamily
  2210.11083}]}.

\bibitem{Fridell:2023gjx}
K.~Fridell, R.~Kitano, and R.~Takai, ``{Lepton flavor physics at
  \ensuremath{\mu}$^{+}$\ensuremath{\mu}$^{+}$ colliders},''
  \href{http://dx.doi.org/10.1007/JHEP06(2023)086}{{\em JHEP} {\bfseries 06}
  (2023) 086}, \href{http://arxiv.org/abs/2304.14020}{[{\ttfamily
  2304.14020}]}.

\bibitem{Bossi:2020yne}
F.~Bossi and P.~Ciafaloni, ``{Lepton Flavor Violation at muon-electron
  colliders},'' \href{http://dx.doi.org/10.1007/JHEP10(2020)033}{{\em JHEP}
  {\bfseries 10} (2020) 033},
  \href{http://arxiv.org/abs/2003.03997}{[{\ttfamily 2003.03997}]}.

\bibitem{Yang:2023ojm}
J.-L. Yang, C.-H. Chang, and T.-F. Feng, ``{Leptonic di-flavor and di-number
  violation processes at high energy colliders},''
  \href{http://dx.doi.org/10.1088/1674-1137/ad17b0}{{\em Chin. Phys. C}
  {\bfseries 48} no.~4, (2024) 043101},
  \href{http://arxiv.org/abs/2302.13247}{[{\ttfamily 2302.13247}]}.

\bibitem{Lichtenstein:2023iut}
G.~Lichtenstein, M.~A. Schmidt, G.~Valencia, and R.~R. Volkas,
  ``{Complementarity of \ensuremath{\mu}TRISTAN and Belle II in searches for
  charged-lepton flavour violation},''
  \href{http://dx.doi.org/10.1016/j.physletb.2023.138144}{{\em Phys. Lett. B}
  {\bfseries 845} (2023) 138144},
  \href{http://arxiv.org/abs/2307.11369}{[{\ttfamily 2307.11369}]}.

\bibitem{Dev:2023nha}
P.~S.~B. Dev, J.~Heeck, and A.~Thapa, ``{Neutrino mass models at $\mu
  $TRISTAN},'' \href{http://dx.doi.org/10.1140/epjc/s10052-024-12496-0}{{\em
  Eur. Phys. J. C} {\bfseries 84} (2024) 148},
  \href{http://arxiv.org/abs/2309.06463}{[{\ttfamily 2309.06463}]}.

\bibitem{Abbiendi:2016xup}
G.~Abbiendi {\em et~al.}, ``{Measuring the leading hadronic contribution to the
  muon g-2 via $\mu e$ scattering},''
  \href{http://dx.doi.org/10.1140/epjc/s10052-017-4633-z}{{\em Eur. Phys. J. C}
  {\bfseries 77} no.~3, (2017) 139},
  \href{http://arxiv.org/abs/1609.08987}{[{\ttfamily 1609.08987}]}.

\bibitem{ParticleDataGroup:2022pth}
{\bfseries Particle Data Group} Collaboration, R.~L. Workman {\em et~al.},
  ``{Review of Particle Physics},''
  \href{http://dx.doi.org/10.1093/ptep/ptac097}{{\em PTEP} {\bfseries 2022}
  (2022) 083C01}.

\bibitem{Bernardi:2022hny}
G.~Bernardi {\em et~al.}, ``{The Future Circular Collider: a Summary for the US
  2021 Snowmass Process},'' \href{http://arxiv.org/abs/2203.06520}{[{\ttfamily
  2203.06520}]}.

\bibitem{Altmannshofer:2022fvz}
W.~Altmannshofer, C.~Caillol, M.~Dam, S.~Xella, and Y.~Zhang, ``{Charged Lepton
  Flavour Violation in Heavy Particle Decays},'' in {\em {Snowmass 2021}}.
\newblock 5, 2022.
\newblock \href{http://arxiv.org/abs/2205.10576}{[{\ttfamily 2205.10576}]}.

\bibitem{Boughezal:2020klp}
R.~Boughezal, C.-Y. Chen, F.~Petriello, and D.~Wiegand, ``{Four-lepton $Z$
  boson decay constraints on the standard model EFT},''
  \href{http://dx.doi.org/10.1103/PhysRevD.103.055015}{{\em Phys. Rev. D}
  {\bfseries 103} (2021) 055015},
  \href{http://arxiv.org/abs/2010.06685}{[{\ttfamily 2010.06685}]}.

\bibitem{CLEO:1995azm}
{\bfseries CLEO} Collaboration, M.~S. Alam {\em et~al.}, ``{Tau decays into
  three charged leptons and two neutrinos},''
  \href{http://dx.doi.org/10.1103/PhysRevLett.76.2637}{{\em Phys. Rev. Lett.}
  {\bfseries 76} (1996) 2637--2641}.

\bibitem{Dicus:1994dt}
D.~A. Dicus and R.~Vega, ``{Standard Model decays of tau into three charged
  leptons},'' \href{http://dx.doi.org/10.1016/0370-2693(94)91389-7}{{\em Phys.
  Lett. B} {\bfseries 338} (1994) 341--348},
  \href{http://arxiv.org/abs/hep-ph/9402262}{[{\ttfamily hep-ph/9402262}]}.

\bibitem{Flores-Tlalpa:2015vga}
A.~Flores-Tlalpa, G.~L\'opez~Castro, and P.~Roig, ``{Five-body leptonic decays
  of muon and tau leptons},''
  \href{http://dx.doi.org/10.1007/JHEP04(2016)185}{{\em JHEP} {\bfseries 04}
  (2016) 185}, \href{http://arxiv.org/abs/1508.01822}{[{\ttfamily
  1508.01822}]}.

\bibitem{Belle-II:2022cgf}
{\bfseries Belle-II} Collaboration, L.~Aggarwal {\em et~al.}, ``{Snowmass White
  Paper: Belle II physics reach and plans for the next decade and beyond},''
  \href{http://arxiv.org/abs/2207.06307}{[{\ttfamily 2207.06307}]}.

\bibitem{Banerjee:2022vdd}
S.~Banerjee, ``{Searches for Lepton Flavor Violation in Tau Decays at Belle
  II},'' \href{http://dx.doi.org/10.3390/universe8090480}{{\em Universe}
  {\bfseries 8} no.~9, (2022) 480},
  \href{http://arxiv.org/abs/2209.11639}{[{\ttfamily 2209.11639}]}.

\end{thebibliography}\endgroup

\end{document}